\documentclass[useAMS,usenatbib,onecolumn]{mn2e}

\pdfoutput=1

\usepackage{amsmath}    % need for subequations
\usepackage{mathtools}  % need for math tools
\usepackage{amsmath}    % need for subequations
\usepackage{amssymb}    % need for symbols
\usepackage{graphicx}   % need for figures
\usepackage{verbatim}   % useful for program listings
\usepackage{color}      % use if color is used in text
\usepackage{subfigure}  % use for side-by-side figures
\usepackage{hyperref}   % use for hypertext links, including those to external documents and URLs
\usepackage{graphicx}   % Used to import the graphics
\usepackage{ifthen}

\newcounter{CDMDone}
\def\CDM{\ifthenelse{\equal{\arabic{CDMDone}}{0}}{cold dark matter (CDM)\setcounter{CDMDone}{1}}{CDM}}

\newcounter{ePSDone}
\def\ePS{\ifthenelse{\equal{\arabic{ePSDone}}{0}}{extended Press-Schechter (ePS)\setcounter{ePSDone}{1}}{ePS}}

\title{Excursion Set Theory for Correlated Random Walks}
\author[Arya Farahi \& Andrew J. Benson]{Arya Farahi$^1$ and Andrew J. Benson$^2$\\
$^1$ Michigan Center for Theoretical Physics, Randall Laboratory of Physics, The University of Michigan, Ann Arbor, MI 48109-1120, USA\\
$^2$ Carnegie Observatories, 813 Santa Barbara Street, Pasadena, CA 91101, USA.}

\begin{document}

\maketitle

\begin{abstract}
We present a new method to compute the first crossing distribution in excursion set theory for the case of correlated random walks. We use a combination of the path integral formalism of Maggiore \& Riotto, and the integral equation solution of Zhang \& Hui, and Benson et al. to find a numerically and convenient algorithm to derive the first crossing distribution. We apply this methodology to the specific case of a Gaussian random density field filtered with a Gaussian smoothing function. By comparing our solutions to results from Monte Carlo calculations of the first crossing distribution we demonstrate that approximately it is in good agreement with exact solution for certain barriers, and at large masses. Our approach is quite general, and can be adapted to other smoothing functions and barrier function, and also to non-Gaussian density fields.
\end{abstract}


\section{Introduction}

The growth of large scale structure in our Universe, and the subsequent formation of galaxies, is a challenging, non-linear problem. Small perturbations in the matter density in the early Universe grow via gravitational instability and eventually collapse to form dark matter halos. Haloes grow through merging with other halos in a hierarchy of ever increasing mass. This hierarchy of dark matter halos provides the cosmic web within which galaxy formation takes places. It is therefore crucial to understand the formation of dark matter halos in order to accurately model the formation of galaxies \citep{Baugh:2006pf}.

N-body simulations have developed into an extremely powerful tool to explore the formation and evolution of dark matter in the highly non-linear regime. However, they remain computationally expensive, making them impractical for surveying dark matter parameter space for example. An alternative approach to computing the formation of dark matter halos was first described by \cite{Press:1973iz}, and later developed extensively by \cite{Bond1991} and \cite{Lacey1993}, using excursion set theory. This \ePS\ methodology provides a highly successful statistical description of the formation histories of dark matter halos. For a recent review of excursion set theory see \cite{Zentner:2006vw}.

The formation of halos has been studied using many different approaches, including numerical approaches such as Monte Carlo (e.g. \citet{Bond1991, 2012Paranjape}), and analytical approaches (e.g. \cite{Bardeen:1986,Bond1991,Maggiore:2009rv, Paranjape:2012pt}). Of these analytic methods, the two most common are peaks theory, \cite{Bardeen:1986}, and the excursion set theory\footnote{Recently, \protect\citep{Paranjape:2012esp} and \protect\cite{Paranjape:2012pt} combined the excursion set and  peak theory methods to construct an excursion set theory for peaks.} \cite{Bond1991}.

The excursion set theory used in the \ePS\ technique follows random walks in the space of over density and variance (of the density field), an associates halo collapse with the first-crossing of a barrier by a random walk. For the case of a constant, or linear barrier and for random walks in which each step is uncorrelated with previous steps, an analytic solution can be found for the distribution of variances (and, therefore, masses) at which halo collapse occurs. This allows the halo mass function to be found directly for the simple case of a Gaussian-random density field and a variance derived using a filter that is sharp in $k$-space. Numerical methods exist to solve the problem numerically in the case of non-linear barriers \citep{Zhang:2005ar,Benson:2012su}, but non-Gaussian density fields and/or different filters, both of which cause successive steps in the random walk to become correlated, are more problematic.

Many attempts have been made to use path integral techniques---a powerful tool for studying stochastic and random processes---to solve problems in excursion set theory (e.g. \cite{Maggiore:2009rv,Maggiore:2009rw,Maggiore:2009rx,Ma:2010ep,Adshead:2012hs,DAloisio:2013}). In this work, we modify the path integral method of \cite{Maggiore:2009rv} and combine it with our modified integral technique for solving the first crossing distribution, which is based upon the work of \cite{Zhang:2005ar} and which we introduced in our previous work, \cite{Benson:2012su}. Through this combination, we develop a new approach to solve the first crossing problem in more general cases, and specifically address the case of a Gaussian filter. In this paper we present our setup for our next work which we will expand this formalism for non-Gaussian random fields.

The remainder of this work is organized as follows. In \S\ref{sec:PathIntegral} we review how the path integral approach is used to find the probability distributions of trajectories and describe how to add a perturbation term representing the effects of the filter. In \S\ref{sec:FirstCrossingProbability} we apply these methods to find the first crossing probability, while in \S\ref{sec:Results} we demonstrate the method using a Gaussian filter. Finally in $\S$\ref{sec:Conculsion} we discuss our results.

We also include three appendices. Appendix~\ref{app:NumericalRecipe} presents our numerical technique for solving the first crossing distribution which we have found to be robust against numerical error. Appendix~\ref{app:MC} describes our method for Monte Carlo simulation of the first crossing distribution, and which is based on the conditional probabilities that we derive through the path integral approach. Finally, Appendix~\ref{app:Claim} shows why it is not possible to find the exact solution to the first crossing problem within this formalism.

%%%%%%%%%%%%%%%%%%%%%%%%%%%%%%%%%%%%%%
%%%%%%%%%%%%%%%%%%%%%%%%%%%%%%%%%%%%%%

\section{Probability Distribution of Trajectory}

\subsection{Introduction}

As shown by \cite{Maggiore:2009rv}, \cite{Maggiore:2009rw}, and \cite{Maggiore:2009rx} (hereafter MR10) in their series of works, path integral methods provide a powerful tool for studying the formation of dark matter halos in the context of excursion set theory. In their works they used this method to address the issues of non-Markovian trajectories, moving barriers, and non-Gaussianity in excursion set theory. Additionally, \cite{Zhang:2005ar} introduced an elegant method for solving the excursion set theory for moving barriers given knowledge of the probability distribution function for over densities as a function of variance.

All of these methods have limitations, which we will discuss further in \S\ref{sec:Conculsion}. In this work we modify the methods of MR10 and \cite{Zhang:2005ar} and combine them to solve the excursion set problem in the general case with greater accuracy. Specifically, our approach differs from that of MR10 as we find the probability distribution function for trajectories without consideration of the barrier, and then combine this with the method of \cite{Zhang:2005ar} to find the first crossing distribution for an arbitrary barrier.

In this section we review how the probability function for a trajectory can be found utilizing path integral techniques, and how one can find mass function for different filters, and we derive all relevant relations.

%%%%%%%%%%%%%%%%%%%%%%%%%%%%%%%%%%%%%%

\subsection{Calculation of the Path Integral} \label{sec:PathIntegral}

We begin by applying the path integral method to the case of a Gaussian random density field with a sharp $k$-space filtering and demonstrate that in this case the well-known analytic solution is recovered. In this work we want to set the stage for more general cases (e.g. non-Gaussian fields), and in future work we plan to comeback to this issue so it worth take sometime and set all pre-requisite tools, for future purposes.

We consider the density perturbation, $\delta(x)$, at point $x$ in the universe, defined as

\begin{equation} \label{eq:densityflactuation}
   \delta(x) = \frac{\rho(x)-\bar{\rho}}{\bar{\rho}}.
\end{equation}

In \ePS\ theory we are interested in a smoothed density field, and so we smooth this density field with some filter function. Assuming a Gaussian-random density field, if the field is smoothed with a sharp $k$-space filter (i.e. a filter that is constant below some critical wavenumber, and zero otherwise) the resulting random walk in $\delta$ as a function of smoothing scale will be a Markovian process. This has a well-known analytic solution for the case of a linear barrier \citep{1998MNRAS.300.1057S,Sheth:2001dp}.

After smoothing, the variance of density field is

\begin{equation} \label{eq:VarianceDef}
   S = \int {\rm d}\ln{k}~\Delta^2(k)~|\tilde{W}^2(k,R)|,
\end{equation}

where $\tilde{W}(k,R)$ is the filter function in Fourier space, $R$ is the smoothing scale, and $\Delta(k)$ is the dimensionless power spectrum. In the path integral approach, we assume that $\delta(S)$ evolves stochastically with ``time-like'' $S$. By definition, $\langle\delta(S)\rangle=0$, and for Gaussian fields the only non-vanishing connected correlator is the two-point correlator, $\langle\delta(S_1)\delta(S_2)\rangle$.

We wish to find the probability distribution of trajectories which start from $S=S_0$, $\delta=\delta_0$ and finish at $S=S_n$, $\delta=\delta_n$. We imagine discretizing our domain, $[S_0,S_n]$ into $n$ equal sized parts. Each step is therefore $\epsilon = \Delta S/n$. The trajectory travels through $S$ and in each step $\delta_i$ takes on a specific value. We will integral over all $\delta_i$'s from $-\infty$ to $\infty$ to find the total probability for the trajectory given its starting and ending points. In the limit $n\rightarrow\infty$ we will recover the continuum solution. We define the probability density in the space of trajectories as

\begin{equation}
   W(\delta_0;\delta_1,\ldots,\delta_n;S_n) \equiv\langle\delta_D(\delta(S_0)-\delta_0)\cdots\delta_D(\delta(S_n)-\delta_n)\rangle,
\end{equation}

where, for clarity, $\delta$ is the density fluctuation field and $\delta_D$ is the Dirac delta function. The probability distribution is found by integrating over all possible paths and has the form

\begin{equation} \label{eq:probabilityDistribution}
   P(\delta_0;S_0;S_1,\ldots,S_{n-1};\delta_n;S_n) = \int_{-\infty}^{+\infty} {\rm d} \delta_1 \ldots \int_{-\infty}^{+\infty} {\rm d} \delta_{n-1} \times  W(\delta_0; \delta_1, \delta_2, \delta_3, \ldots , \delta_{n-1};\delta_n).
\end{equation}

For the case of a Gaussian random density field and a sharp $k$-space filter this probability distribution depends only on the starting and ending points, as we will show below. The function $ W(\delta_0; \delta_1,  \ldots , \delta_{n-1};\delta_n)$ can be expressed in terms of the connected correlators of the theory,

\begin{equation}\label{eq:W}
   W(\delta_0;\delta_1,\ldots,\delta_{n-1};\delta_n) = \int_{-\infty}^{+\infty}\cdots\int_{-\infty}^{+\infty} \left[\frac{{\rm d}\lambda_1}{2 \pi} \cdots \frac{{\rm d}\lambda_n}{2 \pi} e^{i\sum^{n}_{i=1}\lambda_i\delta_i} \langle e^{-i\sum^{n}_{i=1}\lambda_i\delta(S_i)} \rangle \right],
\end{equation}

or more practically,

\begin{equation}
   W(\delta_0;\delta_1,\ldots,\delta_{n-1};\delta_n) = \int_{-\infty}^{+\infty}\cdots\int_{-\infty}^{+\infty} \left[\frac{{\rm d}\lambda_1}{2 \pi} \cdots \frac{{\rm d}\lambda_n}{2 \pi} \exp { \left\{ i\sum\limits_{i=1}^{n}\lambda_i\delta_i + \sum\limits_{p=2}^{\infty} \frac{(-i)^P}{p!} \sum\limits_{i_1=1}^{n} \cdots \sum\limits_{i_p=1}^{n} \lambda_{i_1}  \cdots \lambda_{i_p} \langle\delta_{i_1} \cdots \delta_{i_p}\rangle \right\} } \right].
\end{equation}

Here $\langle\delta_{i_1}\cdots\delta_{i_p}\rangle$ denotes the connected $n$-point correlator. For a Gaussian random field density it is possible to show that the only non-vanishing connected correlator is the two-point correlator, $\langle\delta(S_1)\delta(S_2)\rangle$. Then for the Gaussian case we find, 

\begin{equation} \label{eq:WGaussian}
   W(\delta_0;\delta_1,\ldots,\delta_{n-1};\delta_n) = \int_{-\infty}^{+\infty}\cdots\int_{-\infty}^{+\infty} \left[\frac{{\rm d}\lambda_1}{2 \pi} \cdots \frac{{\rm d}\lambda_n}{2 \pi} \exp{ \left\{ i\sum\limits_{i=1}^{n}\lambda_i\delta_i - \frac{1}{2} \sum\limits_{i,j=1}^{n} \lambda_{i}\lambda_{j} \langle\delta_{i}\delta_{j}\rangle \right\} } \right]
\end{equation}

%%%%%%%%%%%%%%%%%%%%%%%%%%%%%%%%%%%%%%%%%%%%%%%%%%%
\subsection{Gaussian Random Field with Sharp $k$-Space Filtering} \label{sec:Gaussiank-spaceFilter}

It is well-known that for a Gaussian-random density field and sharp $k$-space filter the probability density function for trajectories is

\begin{equation}
   P(\delta_0, S_0=0;\delta_n,S_n) = \frac{1}{\sqrt{2 \pi S_n}} \exp\left\{-\frac{(\delta_n - \delta_0)^2}{2 S_n} \right\}.
\end{equation}

Using eqns.~(\ref{eq:probabilityDistribution}) and (\ref{eq:WGaussian}) we should therefore be able to reproduce this known result. We find

\begin{equation} \label{eq:first2pointCFI}
   P(\delta_0, S_0;\delta_n,S_n) =  \int_{-\infty}^{+\infty} {\rm d} \delta_1 \ldots \int_{-\infty}^{+\infty} {\rm d} \delta_{n-1} 
 \left[ \int_{-\infty}^{+\infty}\cdots\int_{-\infty}^{+\infty} \frac{{\rm d}\lambda_1}{2 \pi} \cdots \frac{{\rm d}\lambda_n}{2 \pi} \exp{ \left\{ i\sum\limits_{i=1}^{n}\lambda_i\delta_i - \frac{1}{2} \sum\limits_{i,j=1}^{n} \lambda_{i}\lambda_{j} \langle\delta_{i}\delta_{j}\rangle \right\} } \right],
\end{equation} 

where $\langle\delta_{i}\delta_{j}\rangle$ is the two point correlation function. MR10 showed that for a Gaussian random field and sharp $k$-space filter the two-point correlator has the form,

\begin{equation} \label{eq:PathIntegralCalculation}
   \langle \delta(S_i)\delta(S_j) \rangle = \min(S_i,S_j) = \epsilon \min(i,j).
\end{equation}  

To solve the integrals in eqn.~(\ref{eq:first2pointCFI}) we will use the following result (see, for example, \citealt{2010CondensedMatterFieldTheory}),

\begin{equation}  \label{eq:IntegralII}
   \int {\rm d}^n x \exp{\left\{ -\frac{1}{2} \sum x_i A_{i,j} x_j + \sum b_i x_i \right\} } = \frac{(2 \pi)^{n/2}}{\sqrt{|A|}} \exp{ \left\{ \frac{1}{2} \sum b_i A_{ij}^{-1} b_j \right\} }.
\end{equation}

We first integrate over the $\lambda$'s in eqn.~(\ref{eq:first2pointCFI}) by setting $b_i \equiv \delta_i$, $A_{ij}\equiv\langle\delta_i\delta_j\rangle$, and $x_i=\lambda_i$ to give\footnote{For a detailed calculation, see MR10.},

\begin{equation} \label{eq:step2}
   P(\delta_0;S_0=0;\delta_n;S_n) = \int_{-\infty}^{+\infty} {\rm d} \delta_1 \ldots \int_{-\infty}^{+\infty} {\rm d} \delta_{n-1} \frac{1}{(2\pi\epsilon)^{n/2}} \exp{ \left\{ - \frac{1}{2\epsilon} \sum\limits^{n-1}_{i=0}(\delta_{i+1} - \delta_i)^2 \right\} }.
\end{equation}

We now employ eqn.~(\ref{eq:first2pointCFI}) again, this time with $x_i=\delta_i$. In this case, the elements of matrix $A$ are $A_{i,i+1}=A_{i,i-1}=-1/\epsilon$ and $A_{i,i}=2/\epsilon$, and the determinant of matrix $A$ is $n\epsilon^{1-n}$. The elements of vector $b$ are: $b_{1} = b_{n-1} = -\epsilon^{-1}$ and the other elements are all zero. Using eqn.~(\ref{eq:IntegralII}) to solve eqn~(\ref{eq:step2}) we find

\begin{equation}
   P(\delta_0;S_0=0;\delta_n;S_n) = \frac{1}{\sqrt{2\pi n\epsilon}} \exp{ \left\{ -\frac{(\delta_n-\delta_0)^2}{2n\epsilon} \right\} }.
\end{equation}

Using the fact that $n\epsilon = S_n$ this reduces so,

\begin{equation} \label{eq:GussiankspaceprobabilityDist}
   P(\delta_0; S_0=0;\delta_n;S_n) = \frac{1}{\sqrt{2\pi S_n}} \exp{ \left\{ - \frac{(\delta_{n} - \delta_0)^2}{2 S_n} \right\} },
\end{equation}

which is independent of our discretization and agrees with the well-known analytic solution.

In the above we have assumed $S_0 = 0$, but in general, one may be interested to have the probability distribution of random walks starting from arbitrary $S_0$. The two-point correlator is unchanged in this case, and so the above calculation holds, but now $n\epsilon = S_n-S_0$, and so

\begin{equation}
   P(\delta_0;S_0;\delta_n;S_n) = \frac{1}{\sqrt{2\pi(S_n-S_0)}} \exp{\left\{ -\frac{ (\delta_n-\delta_0)^2 }{2(S_n-S_0)} \right\} }.
\end{equation}

For solving the excursion set problem it will be useful to calculate the probability of a trajectory which starts at point $S_0=0$, $\delta=0$ and then passes through the point $S=S_i$, $\delta=\delta_i$  before ending at $S=S_n$, $\delta = \delta_n$. We will use this result in section \ref{sec:FirstCrossingProbability}. In this case, we simply  do not integrate over $\delta_i$ in eqn.~(\ref{eq:first2pointCFI}),

\begin{equation}
   P(\delta_0;\delta_i;S_i;\delta_n;S_n) =  \int_{-\infty}^{+\infty} {\rm d} \delta_{1} \ldots \int_{-\infty}^{+\infty} {\rm d} \delta_{i-1}\int_{-\infty}^{+\infty} {\rm d} \delta_{i+1} \ldots \int_{-\infty}^{+\infty} {\rm d} \delta_{n-1} W^{\rm gm}(\delta_0;\delta_1,\cdots,\delta_i,\cdots,\delta_n;S_n).
\end{equation} 

It is then straightforward to show that

\begin{equation} \label{eq:GussiankspaceprobabilityDistII}
   P(\delta_0;\delta_i;S_i;\delta_n;S_n) = \frac{1}{\sqrt{4\pi^2S_i(S_n-S_i)}}\exp\left\{-\frac{(\delta_i-\delta_0)^2}{2S_i}\right\} \exp\left\{-\frac{(\delta_n-\delta_i)^2}{2(S_n-S_i)}\right\}.
\end{equation}

%%%%%%%%%%%%%%%%%%%%%%%%%%%%%%%%%%%%%%%%%%%
\subsection{Gaussian Random Field with Arbitrary Filter} \label{sec:PerturbationCorrection}

In this section we use the prescription introduced in \S\ref{sec:Gaussiank-spaceFilter} to calculate the probability function for trajectories with an arbitrary filter. For an arbitrary filter, considering Gaussian random field, one can solve the necessary integrals analytically. As a result there is no need to go through perturbation theory approach, which MR10 have done in their work.

Without loss of generality will will assume that $\delta_0 = 0$ and $S_0 = 0$. As we are still considering a Gaussian random field all correlations except the two-point correlation will vanish. We then have,

\begin{equation} \label{eq:first2pointCFII}
   P(\delta_0=0;\delta_n;S_n) =  \int_{-\infty}^{+\infty} {\rm d} \delta_1 \ldots \int_{-\infty}^{+\infty} {\rm d} \delta_{n-1} \left[ \int_{-\infty}^{+\infty}\cdots\int_{-\infty}^{+\infty} \frac{{\rm d}\lambda_1}{2 \pi} \cdots \frac{{\rm d}\lambda_n}{2 \pi} \exp{ \left\{ i\sum\limits_{i=1}^{n}\lambda_i\delta_i - \frac{1}{2} \sum\limits_{i,j=1}^{n} \lambda_{i}\lambda_{j} \langle\delta_{i}\delta_{j}\rangle  \right\} } \right] .
\end{equation} 

Using the fact $\int_{-\infty}^{+\infty}{\rm d}\delta~e^{i\lambda\delta} = 2 \pi \delta_D(\lambda)$, where $\delta_D$ is the Dirac delta function, and carrying out all integrals over $\delta_i$'s one get,

\begin{equation}
   P(\delta_0=0;\delta_n;S_n) = \frac{1}{2\pi} \int_{-\infty}^{+\infty}\cdots\int_{-\infty}^{+\infty} {\rm d}\lambda_1 \cdots {\rm d}\lambda_n \prod\limits_{k=1}^{n-1} \delta_D(\lambda_k) \exp{ \left\{ i\lambda_n\delta_n - \frac{1}{2} \sum\limits_{i,j=1}^{n} \lambda_{i}\lambda_{j} \langle\delta_{i}\delta_{j}\rangle  \right\} }  .
\end{equation} 

simply by carrying out all integrals over $\lambda_i$'s one can show,

\begin{equation}
   P(\delta_0=0;\delta_n;S_n) = \frac{1}{\sqrt{2\pi S_n}} \exp{\left\{ -\frac{ \delta_n^2 }{2 S_n} \right\} }.
\end{equation}

Beside the unconditional probability we are interested in conditional probability. We again assume that the trajectory on its way passes through point $S=S_k$, $\delta=\delta_k$. For conditional probability we have,

\begin{equation}
   P(\delta_0=0;\delta_k;S_k;\delta_n;S_n) =  \int_{-\infty}^{+\infty} {\rm d} \delta_{1} \ldots \int_{-\infty}^{+\infty} {\rm d} \delta_{k-1}\int_{-\infty}^{+\infty} {\rm d} \delta_{k+1} \ldots \int_{-\infty}^{+\infty} {\rm d} \delta_{n-1} W^{\rm gm}(\delta_0;\delta_1,\cdots,\delta_k,\cdots,\delta_n;S_n).
\end{equation} 

again by carrying out all integrals over $\delta_i$'s one get,

\begin{align}
   P(\delta_0=0;\delta_k;S_k;\delta_n;S_n) = & \frac{1}{4\pi^2} \int_{-\infty}^{+\infty}{\rm d}\lambda_1 \cdots\int_{-\infty}^{+\infty} {\rm d}\lambda_n \delta_D(\lambda_1) \cdots \delta_D(\lambda_{k-1}) \delta_D(\lambda_{k+1}) \cdots \delta_D(\lambda_{n-1}) \nonumber\\
    &~~~~\times \exp{ \left\{ i\lambda_n\delta_n + i\lambda_k\delta_k  - \frac{1}{2} \sum\limits_{i,j=1}^{n} \lambda_{i}\lambda_{j} \langle\delta_{i}\delta_{j}\rangle  \right\} }.
\end{align} 

using the fact that two-point correlation function is symmetric, $\langle\delta_{i}\delta_{j}\rangle = \langle\delta_{j}\delta_{i}\rangle$ it is straightforward to show

\begin{equation}
   P(\delta_0=0;\delta_k;S_k;\delta_n;S_n) = \frac{1}{4\pi^2} \int_{-\infty}^{+\infty}{\rm d}\lambda_k \int_{-\infty}^{+\infty} {\rm d}\lambda_n  \exp{ \left\{ i\lambda_n\delta_n + i\lambda_k\delta_k + - \frac{1}{2} \lambda_{n}\lambda_{n} \langle\delta_{n}\delta_{n}\rangle - \frac{1}{2} \lambda_{k}\lambda_{k} \langle\delta_{k}\delta_{k}\rangle - \lambda_{k}\lambda_{n} \langle\delta_{k}\delta_{n}\rangle  \right\} }.
\end{equation}

We define $A\equiv\langle\delta_{k}\delta_{k}\rangle$, $B\equiv\langle\delta_{n}\delta_{n}\rangle$, and $C\equiv\langle\delta_{k}\delta_{n}\rangle$. Carrying out above integral get us

\begin{equation} \label{eq:Pmemperturbationterms}
   P(\delta_0=0;\delta_k;S_k;\delta_n;S_n) = \frac{1}{2\pi \sqrt{(AB-C^2)}} \exp\left\{ -\frac{(A\delta_n^2 - 2C\delta_n\delta_k + B\delta_k^2)}{2(AB-C^2)} \right\}.
\end{equation} 

Note that for the case of a sharp $k$-space filter, $A = S_k$, $B=S_n$, and $C=S_k$ then we get the eq.~(\ref{eq:GussiankspaceprobabilityDistII}). For more general cases we again have $A = S_k$ and $B=S_n$, but $C$ now depends on the choice of filter and power spectrum.

This formalism also provides an alternative Monte Carlo approach to excursion set theory which is more time efficient than that of \citet{Bond1991}. We describe this approach, and demonstrate that it gives the same answers as that of \citet{Bond1991} in App.~\ref{app:MC}.

%%%%%%%%%%%%%%%%%%%%%%%%%%%%%%%%%%%%
\subsection{Calculating the Filter Effect} \label{sec:calculationoftheFE}

We define filter effect (hereafter FE) as the deviation of two-point function from that of the two-point function of $k$-space filter, $\Delta(S_i,S_j)=\Delta_{ij}$. As we will see later in section \ref{sec:FirstCrossingProbability} particular advantage of our method compared to its ancestors is that the terms that contribute to the FE are completely independent of the solution. This makes it very easy to develop solutions for alternative filters which can be used in eqn.~(\ref{eq:Pmemperturbationterms}). As we will see, $C$ which is defined in eqn.~(\ref{eq:Pmemperturbationterms}) is $\min{(S_i,S_j)} + \Delta_{ij}$.

The two point correlation function of the sharp $k$-space filter was $\min{(S_i,S_j)}$. Therefore,

\begin{equation} \label{eq:FilterEffectTermGeneral}
   \Delta_{ij} = \langle \delta_i\delta_j \rangle - \min{(S_i,S_j)}.
\end{equation}

One can use the above formula to find the perturbation formula for any filter of interest. Note that, for most cases there will be no analytical solution, and the filter effect must be computed numerically for each point.

To be specific, we will now calculate the FE for a Gaussian filter. For simplicity, we consider a power spectrum of the form $P(k)\propto k^n$ with $n = 1$. Using eqn.~(\ref{eq:twopointcorrelationGuassianFilter1}) and assuming $i<j$ we find

\begin{equation} \label{eq:EFtemGaussianFilter}
   \Delta^{\rm GF}_{ij} = \frac{3S_iS_j - S_i^2 -2\sqrt{(S_i^3S_j)}}{S_i + S_j + 2\sqrt{S_iS_j}},
\end{equation}

where GF stands for ``Gaussian filter''. Note that $\Delta^{\rm GF}_{ij}$ in eqn.~(\ref{eq:EFtemGaussianFilter}) is not symmetric in its arguments as we have explicitly assumed that $i < j$.

\begin{figure}
 \centering
 \includegraphics[width=8.0cm]{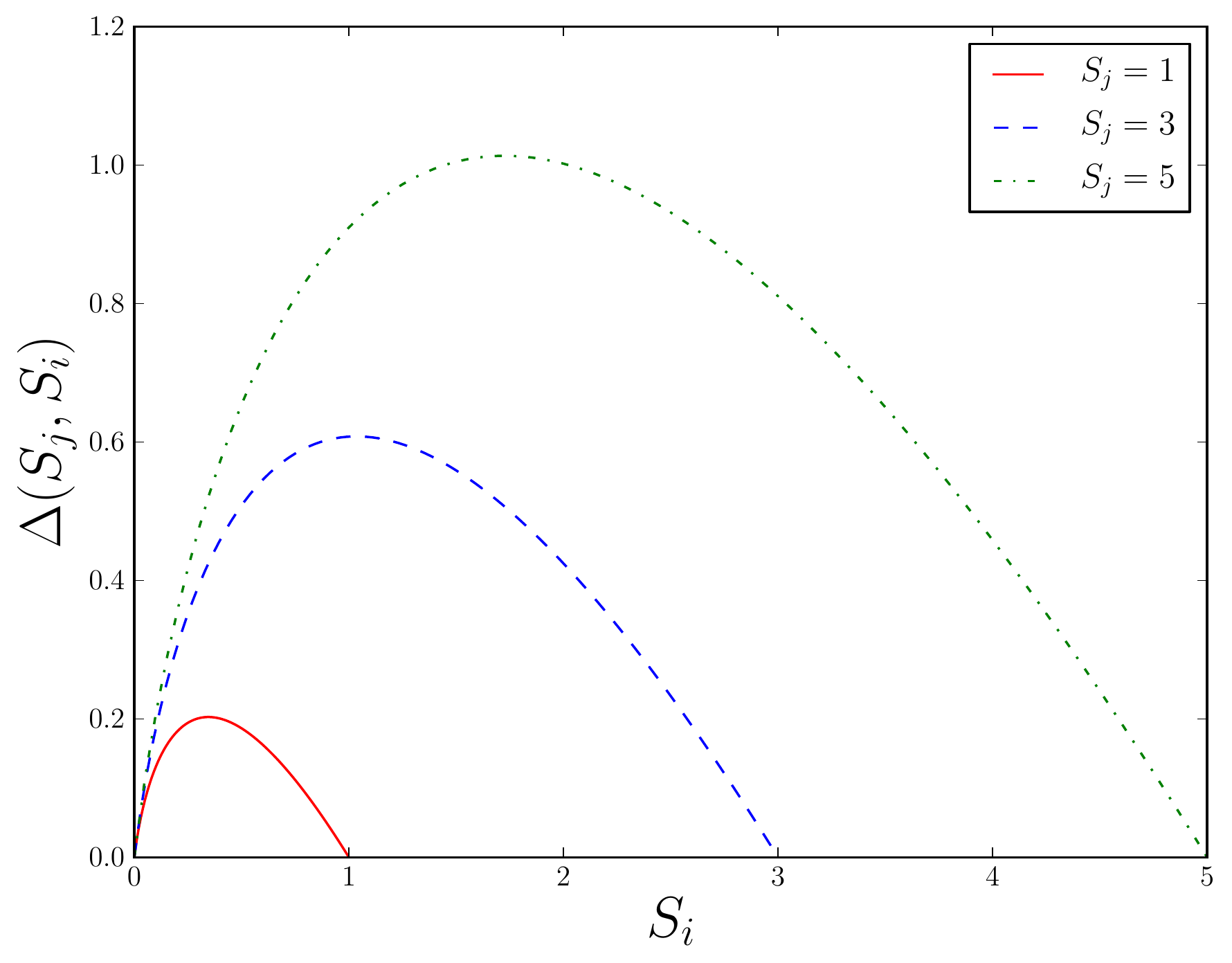}
 \caption{The FE for a Gaussian filter as given by eqn.~(\ref{eq:EFtemGaussianFilter}). Red (solid line), blue (dashed line), and green (dot-dashed line) lines show the FE for $S_j=1,~3,~5$ respectively.}
 \label{fig:FEComparison}
\end{figure}

In Fig.~\ref{fig:FEComparison} we show the FE at three points. As one can see in Fig.~(\ref{fig:FEComparison}), as the smoothing scale, $S$, increases the FE become more important.

%%%%%%%%%%%%%%%%%%%%%%%%%%%%%%%%%%%%
%%%%%%%%%%%%%%%%%%%%%%%%%%%%%%%%%%%%
\section{First Crossing Probability} \label{sec:FirstCrossingProbability}

\citet{Zhang:2005ar} introduced an elegant method to find the first crossing probability in excursion set theory for arbitrary barriers. In our previous work \citep{Benson:2012su} we modified this method to obtain more accurate answers for moving barriers. In this section we will incorporate the filter effect into this integral formulation. The method of \citet{Zhang:2005ar} begins by considering the integral form of the first crossing problem, which simply states that the total probability at any variance $S$ is the sum of the probability that a trajectory has crossed the barrier at some smaller $S$, and the probability that the trajectory is currently below the barrier having never crossed at any smaller $S$:

\begin{equation} \label{eq:Zhang:2005arformulation}
   1 = \int_{0}^{S_n}f(S^\prime){\rm d}S^\prime + \int^{B_a(S_n)}_{-\infty}P_{\rm B}(\delta,S_n){\rm d}\delta
\end{equation}

where $P_{\rm B}(\delta,S_n)$ is the distribution of $\delta$ at $S_n$ \emph{accounting for the absorbing barrer} and is given by

\begin{equation} \label{eq:Zhang:2005arformulationP0}
   P_{\rm B}(\delta_n,S_n) = P(\delta_n,S_n) - \int^{S_n}_{0} G(S_n,\delta_n,S^\prime){\rm d}S^\prime.
\end{equation}

In this formulation, eqns.~(\ref{eq:Zhang:2005arformulation}) and (\ref{eq:Zhang:2005arformulationP0}), $G(S_n,\delta_n,S^\prime){\rm d}S^\prime$ is the probability for a trajectory which crosses the barrier for the first time between $S^\prime$ and $S^\prime+{\rm d}S^\prime$ and approaches the point $(\delta_n,S_n)$. It is calculated by multiplying the fraction of trajectories which cross the barrier between $S^\prime$ and $S^\prime+{\rm d}S^\prime$ for the first time, $f(S^\prime){\rm d}S^\prime$, by the probability, $P(\delta_0;S_0;S^\prime,B_a(S^\prime);\delta_n;S_n)$, of a trajectory which starts from point $(S_0=0,\delta=0)$, crosses the barrier at point $(S=S^\prime,\delta=B_a(S^\prime))$, and finally ends at point $(S=S_n,\delta = \delta_n)$. Here, $B_a(S)$, is the moving barrier. In general one can calculate the second term of equation \ref{eq:Zhang:2005arformulationP0} using,

\begin{equation}
   G(S_n,\delta_n,S^\prime) \equiv f(S^\prime)\frac{P(\delta_0;S_0;S^\prime,B_a(S^\prime);\delta_n;S_n)}{P(\delta_0;S_0;S',B_a(S^\prime))} ,
\end{equation}

where $P(\delta_0;S_0;S^\prime,B_a(S^\prime))$ is probability of trajectory which starts from point $(S=S_0,\delta=\delta_0)$ and ends at point $(S=S^\prime,\delta=B_a(S^\prime))$. In the case of a Gaussian-random field and a sharp $k$-space filter one can divide eqn.~(\ref{eq:GussiankspaceprobabilityDistII}) by eqn.~(\ref{eq:GussiankspaceprobabilityDist}) and obtain

\begin{equation} \label{eq:IntegralEquationCondition}
   G(S,\delta,S^\prime,B_a(S^\prime)) = f(S^\prime) \frac{1}{\sqrt{2\pi(S-S^\prime)}} \exp{\left\{-\frac{(\delta - B_a(S^\prime))^2}{2(S-S^\prime)}\right\} } = f(S^\prime)P_0(\delta-B_a(S^\prime),S-S^\prime).
\end{equation}

For the case of a Gaussian-random field and a generic filter one can use the exact solution of eqn.~(\ref{eq:Pmemperturbationterms}) to obtain\footnote{Note that, in the general case, one should consider not the fraction of trajectories passing through point $(B_a[S],S)$ which end at point $(\delta_n,S_n)$, but the fraction of trajectories which pass through point $(B_a[S],S)$ \emph{having never crossed the barrier for smaller $S$} and which then end at point $(\delta_n,S_n)$. For a Gaussian random field with a sharp $k$-space filter, these two are equivalent, as the path of any given trajectory for $S^\prime > S$ is independent of its path for $S^\prime < S$. We thank Ravi Sheth for pointing out this issue.\label{fn3}},

\begin{equation} \label{eq:IntegralEquationConditionGeneral}
   G(S,\delta,S^\prime,B_a(S^\prime)) = f(S^\prime) \frac{\sqrt{A}}{\sqrt{2\pi(AB-C^2)}} \exp\left\{ -\frac{(A~\delta-C~B_a(S^{\prime}))^2}{2A(AB-C^2)} \right\} .
\end{equation}

with $A\equiv\langle\delta^{\prime}\delta^{\prime}\rangle=S^{\prime}$, $B\equiv\langle\delta \delta\rangle=S$, and $C\equiv\langle\delta^{\prime}\delta\rangle$, and finally $B_a(S^{\prime})$ is the barrier function. As we have shown in \cite{Benson:2012su}, to discretize this equation and find the first crossing probability, $f(S)$, the next step is to find the following integral,

\begin{align} \label{eq:FilterCorrectionTermIntegral}
   \int^{B_a(S)}_{-\infty} G(S,\delta,S^\prime,B_a(S^\prime)) {\rm d}\delta = \frac{f(S^{\prime})}{2} \left[1 + \hbox{erf}\left\{\frac{A~B_a(S)-C~B_a(S^\prime)}{\sqrt{2A(AB-C^2)}}\right\} \right].
\end{align}

in a case of a sharp $k$-space filter it simplifies to,

\begin{align}
   \int^{B_a(S)}_{-\infty} G(S,\delta,S^\prime,B_a(S^\prime)) {\rm d}\delta = \frac{f(S^{\prime})}{2} \left[1 + \hbox{erf}\left\{\frac{B_a(S)-B_a(S^\prime)}{\sqrt{2(S-S^\prime)}}\right\} \right].
\end{align}

We can now discretize eqn.~(\ref{eq:Zhang:2005arformulation}) and solve for $f(S)$. In our previous work we showed in detail how one can discretize this equation. Briefly, we have

\begin{equation} \label{eq:IntegralMethodWithPertTerm}
   1 = \int_{0}^{S}f(S^\prime){\rm d}S^\prime + \hbox{erf}\left\{\frac{B_a(S)}{\sqrt{2S}}\right\} - \int_{0}^{S}f(S^\prime)\times\hbox{erf}\left\{\frac{A~B_a(S)-C~B_a(S^\prime)}{\sqrt{2A(AB-C^2)}}\right\}~{\rm d}S^\prime.
\end{equation}

We assume that $\Delta S \equiv S_{\max}/N$. For eqn.~(\ref{eq:Zhang:2005arformulation}) we find

\begin{align}
   1 =& \sum\limits_{i=0}^{j-1}  \frac{f(S_i) + f(S_{i+1})}{2}\Delta S + \hbox{erf}{\left\{\frac{B_a(S_j)}{\sqrt{2(S_j)}}\right\}} \nonumber\\
      &~ - \sum\limits_{i=0}^{j-1} \left[ f(S_i)\times \hbox{erf}\left\{\frac{A_i~B_a(S_j)-C_{i,j}~B_a(S_i)}{\sqrt{2A_i(A_iB_j-C_{i,j}^2)}}\right\} + f(S_{i+1})\times\hbox{erf}\left\{\frac{A_{i+1}~B_a(S_j)-C_{i+1,j}~B_a(S^\prime)}{\sqrt{2A_{i+1}(A_{i+1}B_j-C_{i+1,j}^2)}}\right\}\right] {\Delta S\over 2}.
\end{align}

Since $f(0) = 0$ we obtain,

\begin{align} \label{eq:NumericalDiscritizationFinal}
   \left[1 - \hbox{erf}\left\{\frac{A_j~B_a(S_j)-C_{j,j}~B_a(S_j)}{\sqrt{2A_j(A_jB_j-C_{j,j}^2)}}\right\} \right]\frac{\Delta S}{2}f(S_j) = & 1 - \hbox{erf}{\left\{\frac{B(S_j)}{\sqrt{2(S_j)}}\right\}} \nonumber\\
    &~~~ - \sum\limits_{i=1}^{j-1} f(S_i)\left(1 - \hbox{erf}\left\{\frac{A_i~B_a(S_j)-C_{i,j}~B_a(S_i)}{\sqrt{2A_i(A_iB_j-C_{i,j}^2)}}\right\}\right)\Delta S.
\end{align}

For all barriers of interest $\hbox{erf}\left\{\frac{A_j~B_a(S_j)-C_{j,j}~B_a(S_j)}{\sqrt{2A_j(A_jB_j-C_{j,j}^2)}}\right\}=0$, such that

\begin{equation} 
   f(S_j) = \left[ 1 - \hbox{erf}{\left\{\frac{B_a(S_j)}{\sqrt{2(S_j)}}\right\}} - \sum\limits_{i=1}^{j-1} f(S_i)\left(1 - \hbox{erf}\left\{\frac{A_i~B_a(S_j)-C_{i,j}~B_a(S_i)}{\sqrt{2A_i(A_iB_j-C_{i,j}^2)}}\right\} \right)\Delta S \right] \frac{2}{\Delta S}.
\end{equation}

Note that in above equation $A_i$, $B_j$, and $C_{i,j}$ are just function of smoothing scale, $S$. One can check that in case of sharp $k$-space filter above formula simplifies to what \citep{Benson:2012su} derived in their work. We address numerical issues in solving for $f(S_j)$ in Appendix~\ref{app:NumericalRecipe}.

%%%%%%%%%%%%%%%%%%%%%%%%%%%%%%%%%%%%%%
\subsection{Gaussian Random Field with Gaussian filter} \label{sec:GaussianGaussianFilter}

So far, our calculations are independent of the choice of filter. In this section we compute the first crossing distribution for a Gaussian filter applied to a Gaussian-random field. For a Gaussian-random field both the unconditional and conditional probabilities should also be Gaussian, independent of the choice of filter. We use this case to illustrate that the path integral formalism gives the expected answer. Of course, the true power of the path integral formalism will become relevant when applied to non-Gaussian fields. The Gaussian filter is convenient as its two point correlation function can be calculated analytically and, unlike the sharp $k$-space filter, its smoothing region in configuration space is well defined and localized. In this case, the variance, $S$, is defined as

\begin{equation}
   S = \langle \delta^2(R) \rangle = \frac{1}{2\pi^2} \int k^2~P(k)~\tilde{W}^2(k,R)~{\rm d}k,
\end{equation}

where

\begin{equation}
 \tilde{W}(k,R) = \exp\left\{-\frac{R^2k^2}{2}\right\}
\end{equation} 

is the Fourier transform of our filter which in real space has the form

\begin{equation}
   W(r,R) = \frac{1}{(2\pi R^2)^{3/2}}\exp\left\{-\frac{r^2}{2R^2}\right\}.
\end{equation}

For a scale-free power spectrum, $P(k) = Ak^n$,

\begin{equation}
   S = \langle \delta^2(R) \rangle = \frac{A}{4\pi^2} \Gamma\left(\frac{3+n}{2}\right) R^{-3-n}~,
\end{equation}

and two-point correlation function is

\begin{equation}
   \langle \delta(R_1)\delta(R_2) \rangle = \frac{A}{4\pi^2}\Gamma\left(\frac{3+n}{2}\right)\left(\frac{R_1^2}{2}+\frac{R_2^2}{2}\right)^{\frac{-3-n}{2}}~,
\end{equation}

where here $\Gamma$ is the usual Gamma function. For $n = 1$ therefore, we find the following relation between the variance and the two-point correlation function:

\begin{equation} \label{eq:twopointcorrelationGuassianFilter1}
   \langle\delta_i\delta_j\rangle = \frac{4S_iS_j}{S_i+S_j+2\sqrt{S_iS_j}}.
\end{equation}

For general exponent $n$ this becomes:

\begin{equation}
   \langle\delta_i\delta_j\rangle = \left( \frac{S_i^{\frac{-2}{3+n}} + S_j^{\frac{-2}{3+n}} }{2} \right)^{-\frac{3+n}{2}}.
\end{equation}

Setting $i=j$ gives us the variance of point $S_i$, $\langle\delta_i\delta_j\rangle =S_i$, consistent with that found for a sharp $k$-space filter for which $\langle\delta_i\delta_j\rangle=\min(S_i,S_j)$. Note that for a Gaussian filter one can derive the two point correlation function analytically. For more complicated cases there is no analytical formula of two point function so the two point function must be found numerically.

%%%%%%%%%%%%%%%%%%%%%%%%%%%%%%%
%%%%%%%%%%%%%%%%%%%%%%%%%%%%%%%

\section{Results} \label{sec:Results}

We can now use our results to compute the first crossing distribution for a Gaussian filter. MR10 also employed the path integral method account for the filter effect as perturbation term. They used an approximate form for the two-point correlation function for both a top-hat filter in real space and a Gaussian filter. Specifically, they proposed the following function:

\begin{equation} \label{eq:MR10eq} 
   f(S) = \frac{1-\kappa}{\sqrt{2\pi}}\frac{\delta_c}{S^{3/2}}\exp{\left\{-\frac{\delta^2_c}{2S}\right\}} + \frac{\kappa}{2\sqrt{2\pi}}\frac{\delta_c}{S^{3/2}}\Gamma\left(0,\frac{\delta_c^2}{2S}\right),
\end{equation}

where $\Gamma(0,z)$ is the incomplete Gamma function, $\delta_{\rm c}\approx 1.686$ is the barrier for spherical collapse. MR10 computed the value of $\kappa$ for top-hat and Gaussian filters, and a $\Lambda$CDM power spectrum, finding that $\kappa\approx0.4592$ and $0.35$ respectively with a weak dependence on scale. Because MR10 computed $\kappa$ for a $\Lambda$CDM power spectrum it is not directly comparable to the power-law cases we consider here. Furthermore, for case of power-law, we have compared the form of our expression for $\Delta_{ij}$ and find that it is not well approximated by MR10's fitting function for any value of $\kappa$. Therefore, we instead compare our solutions to the results of Monte Carlo simulations of the first crossing problem.

\cite{Peacock:1990zz} used an alternative method to find the first crossing probability distribution for correlated random walks, proposing an approximate formula for the Gaussian filter. Later \cite{Paranjape:2011wa} proposed another approximate formula and argued that their formula worked well for a large range of $S$. More recently, MS12 used this method to propose an improved approximation, for generic filters as well as moving barriers. 

\begin{figure}
 \centering
 \begin{tabular}{cc}
 \includegraphics[width=8.5cm]{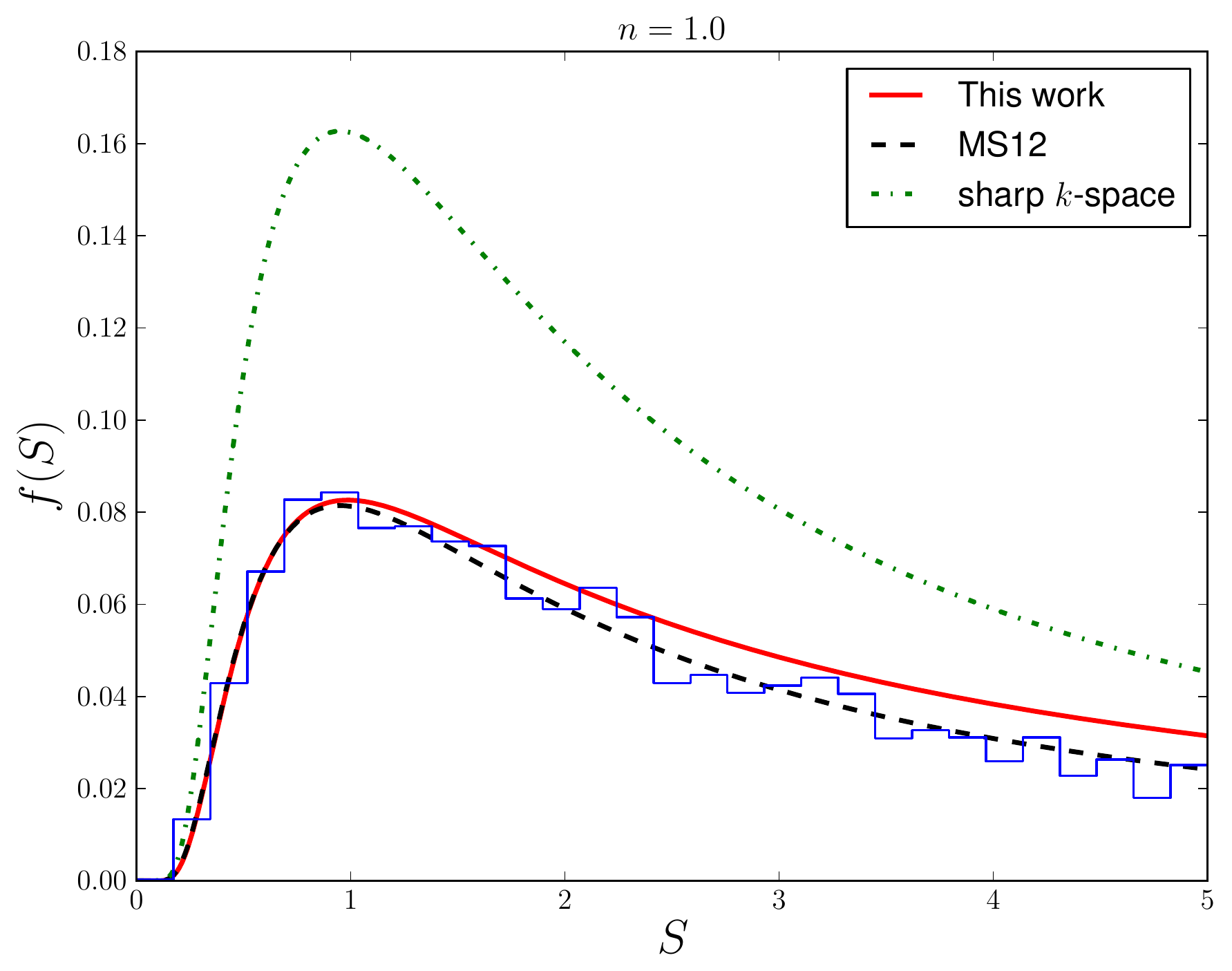} &
 \includegraphics[width=8.5cm]{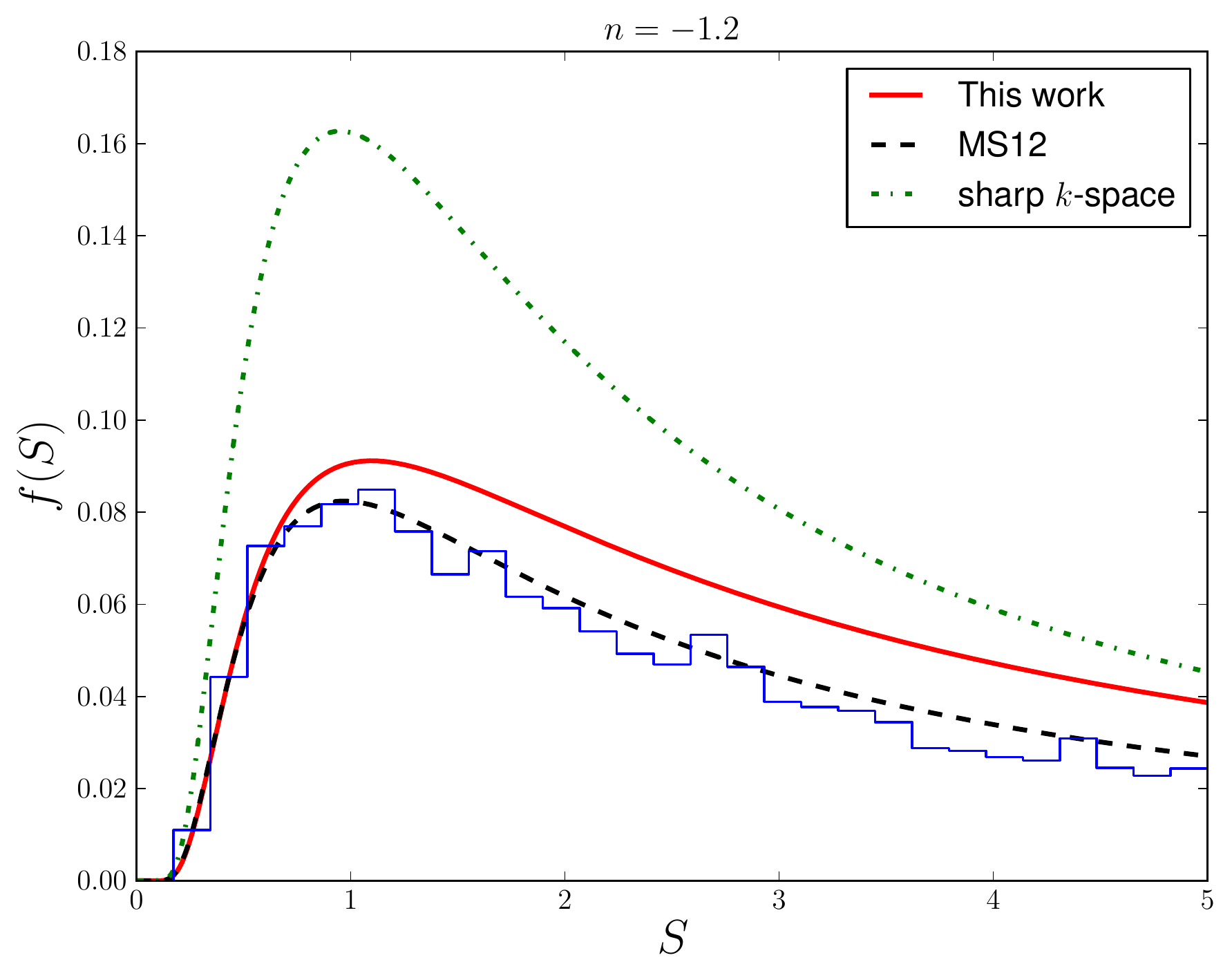}
 \end{tabular}
 \caption{The halo mass function $f(S)$ as computed using the numerical solution of equ(\ref {eq:FilterCorrectionTermIntegral}) and approximation solution of MS12, for a Gaussian-random density field, smoothed with a Gaussian filter, and for two different power spectral indices: $n=+1.0$ (left panel) and $n=-1.2$ (right panel). The dark blue line shows results from a Monte Carlo calculation, while red line (solid line) shows results from this work. Black line (dashed line) show the approximation of MS12. For reference, the green (dot-dashed line) line shows the exact result for a sharp $k$-space filter.}
 \label{fig:ResultsMS12} 
\end{figure}

For comparison we use power-low power spectrum with $n=+1.0$ (which approximates the power spectrum of the density field immediately after inflation and on very large scales today) and $n=-1.2$ which approximates the $\Lambda$CDM power spectrum on the scale of galaxies today. Fig.~\ref{fig:ResultsMS12} compares results from this work with those of MS12, and a Monte Carlo calculation of trajectories\footnote{We cannot compare with the results of MR10 in this case. Their mass function is expressed in terms of a parameter, $\kappa$, which parameterizes the perturbation, $\Delta_{ij}$. We find that the parameterization used by MR10---derived for a $\Lambda$CDM power spectrum---is not a good match to the perturbation derived in this work for any value of $\kappa$. Note that MR10 used transfer function for calculating  $\kappa$ but in this work just the power-law is considered. For similar reasons, the work of \cite{Corasaniti:2011dr} and \cite{Achitouv:2011sq} are also not comparable to our current examples.}. We use \cite{Robertson:2009A}'s prescription (see also \citealt{Bond1991}) for Monte Carlo simulation of random walks. Results of \cite{Robertson:2009A}'s prescription is similar to results of prescription explained in appendix \ref{app:MC}. Results are shown for two different values of the power spectrum index, $n$. For $n=+1.0$ (left panel) and $n=-1.2$ (right panel) our results are in very good agreement with the Monte Carlo and MS12's results for small $S$, but diverge from the exact solution as $S$ increasese. This divergence occurs because we cannot fully account for the correlated nature of the walks (see footnote~\ref{fn3}).

\begin{figure}
 \centering
 \begin{tabular}{cc}
 \includegraphics[width=8.5cm]{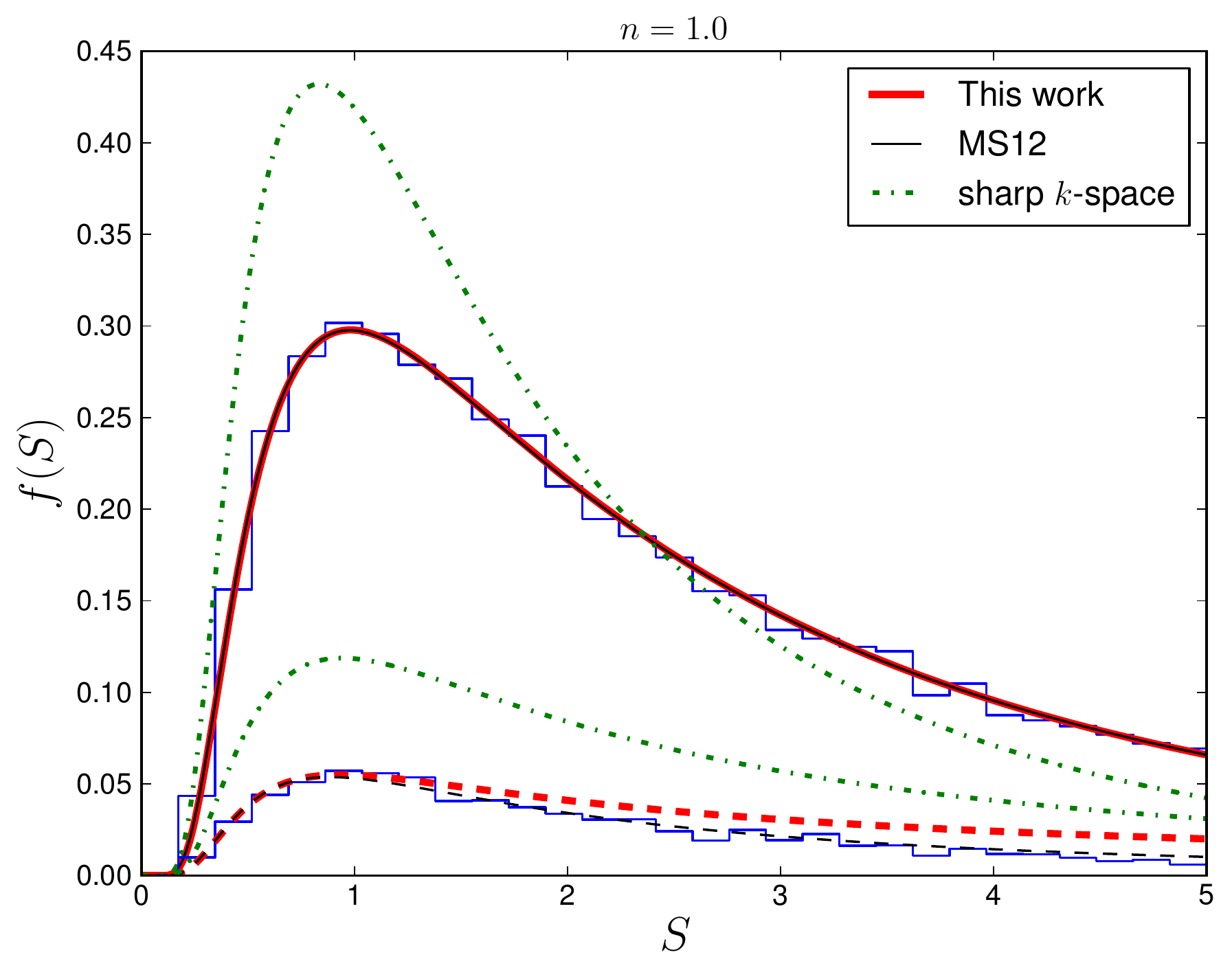} &
 \includegraphics[width=8.5cm]{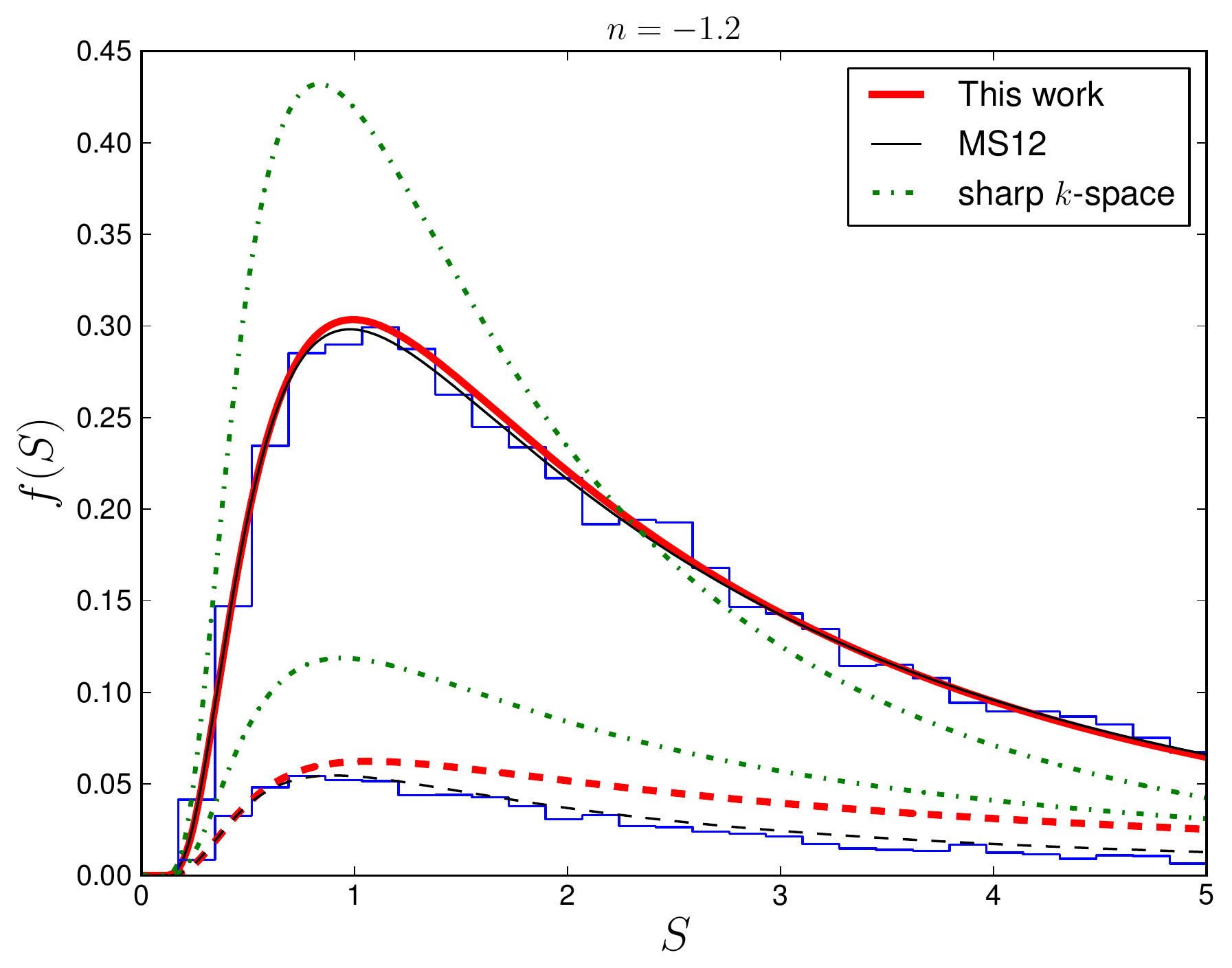}
 \end{tabular}
 \caption{The halo mass function $f(S)$ as computed using different approximations, for a Gaussian-random density field, smoothed with a Gaussian filter, and for two different power spectral indices: $n=+1.0$ (left panel) and $n=-1.2$ (right panel), for the case of a linear moving barrier of the form $B(S) = \delta_{\rm c}(1+\alpha S/\delta_{\rm c}^2)$. Dashed lines show results for $\alpha=+0.3$ while solid lines show results for $\alpha=-1.2$. The dark blue lines show results from a Monte Carlo calculation, while red lines (thick lines) show results from this work. Black lines (thin lines) show the approximation of MS12. For reference, the green lines (dot-dashed lines) shows the exact result for a sharp $k$-space filter.}
 \label{fig:ResultsMovingBarrier} 
\end{figure}

Our method applies equally well to constant and non-constant (a.k.a. ``moving'') barriers. Figure~\ref{fig:ResultsMovingBarrier} compares results from this work with those of MS12, and a Monte Carlo calculation of trajectories for a linear moving barrier with the form of,

\begin{equation}
  B(S) = \delta_{\rm c}\left(1+\frac{\alpha S}{\delta_{\rm c}^2}\right).
\end{equation}

The figure shows the result for two choices of $\alpha$, $+0.3$ (dashed lines) and $-1.2$ (solid lines), and two different power spectra, $n=+1.0$ (left panel) and $n=-1.2$ (right panel). For negative slopes, $\alpha < 0$, the approximation of MS12, this work, and Monte Carlo calculations are in close agreement with each other. For positive slopes, $\alpha > 0$, all methods agree for small $S$, but as $S$ increases our method diverges from the Monte Carlo, again due to our inability to fully accounted for the correlated nature of random walks.

%%%%%%%%%%%%%%%%%%%%%%%%%%%%%%%%%%%%%%%%%%%%%%%%%%%%%%
%%%%%%%%%%%%%%%%%%%%%%%%%%%%%%%%%%%%%%%%%%%%%%%%%%%%%%
\section{Conclusion and Discussion} \label{sec:Conculsion}

Many attempts have been made to account for correlated random walks in excursion set theory, whether induced by the filter effect or by intrinsic correlations in the density field. For example, \cite{Peacock:1990zz} computed approximate solutions for correlated random walks, specifically for a Gaussian filter function. This approach was later improved \citep{Paranjape:2011ak,Paranjape:2011wa,Musso:2012qk,Musso:2012ch} to work with moving barriers.  While direct Monte Carlo simulation can account for these correlations it is inefficient and therefore impractical for application within semi-analytical models which use \ePS\ theory to construct merging histories for dark matter halos. MR10 employed a path integral method to include the filter effect as a perturbation term. They used an approximate function for the two-point correlation function arising from both real-space top-hat, and Gaussian filters. In their later works they have proposed a series solution for moving barriers with sharp $k$-space filters. It is difficult to include both the filter effect and a moving barrier in the approach of MR10. Later works \citep{Corasaniti:2011dr} improved MR10's method by considering a linear barrier in the perturbative path integral method, but still are not applicable to an arbitrary moving barrier, and also their method is based on perturbative expansion. 

The solution for arbitrary moving barriers with uncorrelated random walks is well understood and has been studied for many years \citep{Zentner:2006vw}. In particular, the integral method of \cite{Zhang:2005ar}, and its numerically more robust modification \citep{Benson:2012su}, work well in this case. In this work we have combined the power of the path-integral approach to excursion set theory with the integral-method solution for arbitrary barriers and develop proposed an approximate solution to the first crossing problem applicable to arbitrary filters and barriers. Our approach utilizes the path integral formalism of MR10, but uses it to find the unconditional probability $P_0(\delta,S)$---the probability that a random walk reaches the point $\delta$ at variance $S$---and the conditional probability $P(\delta_c,S_c;\delta,S)$---the probability that a random walk cross the barrier at point $S$ then reaches the point $\delta$ at variance $S$--- considering all possible walks and weighting them equally when constructing the ensemble average. Note that we specifically \emph{do not} use the path integral formalism to compute the probability $P(\delta,S)$ in the presence of the absorbing barrier. In this formalism, the conditional probability is responsible for accounting for correlations between steps in the random walk. Given these unconditional and conditional probabilities, the barrier is added to the problem and used to compute the first crossing distribution using the method of \cite{Zhang:2005ar}.

Our approach is an approximate solution to this problem applicable to all possible barrier shapes and filter functions. In the limit of constant barrier and with a Gaussian filter our method matches the results of MS12 and Monte Carlo calculations results very well for small $S$. This can be clearly seen in Figure~\ref{fig:ResultsMovingBarrier} in which we show results for $n=-1.2$ and $n=+1.0$. We also compared our solution with that of MS12 for a linear moving barrier with the form $B(S) = \delta_{\rm c}(1+\alpha S/\delta_{\rm c}^2)$ for two choice of $\alpha$, $\alpha=+0.3$ and $\alpha=-1.0$. As can be seen in Figure~\ref{fig:ResultsMS12}, our method is in excellent agreement with Monte Carlo calculations for small $S$'s for $\alpha>0$, and for all ranges in $S$ for moving barriers with $\alpha < 0$, although in the case of a Gaussian filter the MS12 approximation works significantly better for the constant and increasing barriers over a wide range of $S$.

In this work we showed that approximating the conditional probability function (i.e. the probability for a trajectory to arrive at a given end point, $(\delta_n,S_n)$, under the condition that it passed through some intermediate point, $(B_a(S),S)$) by simply considering the fraction of trajectories passing through point $(B_a(S),S)$ which end at point $(\delta_n,S_n)$ leads to inaccuracies in our method (which involves no other approximations). Finding the exact analytical result would require that we account for only those trajectories which pass through point $(B_a(S),S)$ having \emph{never} crossed the barrier for smaller $S$. In Appendix~\ref{app:Claim} we outline why we have been unable to solve this problem.

Currently, we have used these methods to derive the unconditional and conditional probability distributions (required to solve the first crossing problem) for the simple case of a Gaussian random field---demonstrating that this approach correctly recovers the expected Gaussian distributions in this case. In a future work, we intend to exploit this flexibility of our method to examine models in which the density field is non-Gaussian, and we aim to use the setup intorduce here. Additionally, we note that our approach currently gives equal weight to each possible random walk when constructing the ensemble average. Dark matter halos may, in fact, correspond to a special subset of all walks, e.g. those corresponding to peaks in the density field \citep{Paranjape:2011ak,Paranjape:2012esp}, which would require different weights to be assigned to random walks in the ensemble averaging. We plan to revisit this issue.

\section*{Acknowledgments}
We thank Ravi Sheth, Brant Robertson, and Antonio Riotto for helpful discussions, and the anonymous referee for recommendations which improved this work.

\bibliographystyle{mn2e}
\bibliography{mybib}

\appendix

\section{Numerical Algorithm}\label{app:NumericalRecipe}

In our previous work \citep{Benson:2012su} and here in \S\ref{sec:FirstCrossingProbability} we employ a first order finite difference method to discretize and solve the integral equation~(\ref{eq:IntegralMethodWithPertTerm}). The solution to this equation contains high-order cancellations which can make it numerically challenging to solve. We have found that for a sharp $k$-space filter this method is robust for all $S$ and for arbitrary barriers providing that care is taken in the numerical evaluation. However, we find that filter functions can induce numerical inaccuracies that lead to fluctuation and divergence of the solution for $S>1$. To circumvent this problem we proceed as follows.

We begin by rewriting eqn.~(\ref{eq:NumericalDiscritizationFinal}) for sharp $k$-space filter:

\begin{align} \label{eq:NumericalDiscritizationFinalApp}
   1 - \hbox{erf}{\left\{\frac{B(S_j)}{\sqrt{2(S_j)}}\right\}} = \left[1 - \hbox{erf}{\left\{ \frac{B(S_j)-B(S_j)}{ \sqrt{2(S_j-S_j)} } \right\} } \right]\frac{\Delta S}{2}f(S_j) + \sum\limits_{i=1}^{j-1} f(S_i)\left(1 - \hbox{erf}{\left\{\frac{B(S_j)-B(S_i)}{\sqrt{2(S_j-S_i)}}\right\}} \right)\Delta S.
\end{align}

We now adjust the weight given to the first term on the right-hand side of eqn.~(\ref{eq:NumericalDiscritizationFinal}) from $\Delta S / 2$ to $\Delta S/\alpha$. Solving for $f(S_j)$ gives

\begin{equation} 
   f(S_j) = \left[ 1 - \hbox{erf}{\left\{\frac{B(S_j)}{\sqrt{2(S_j)}}\right\}} - \sum\limits_{i=1}^{j-1} f(S_i)\left(1 - \hbox{erf}{\left\{\frac{B(S_j)-B(S_i)}{\sqrt{2(S_j-S_i)}}\right\}} \right)\Delta S \right] \frac{\alpha}{\Delta S}.
\end{equation}

Note that $f(S_j)$ is proportional to $\alpha$. However, all $f(S_{i<j})$ also depend on $\alpha$. In particular, if we make $\alpha$ slightly smaller than $2$ we cause successive $f(S_j)$ to oscillate around the true solution, effectively damping the numerical noise which otherwise causes divergence.

\begin{figure}
 \centering
 \includegraphics[width=12.0cm]{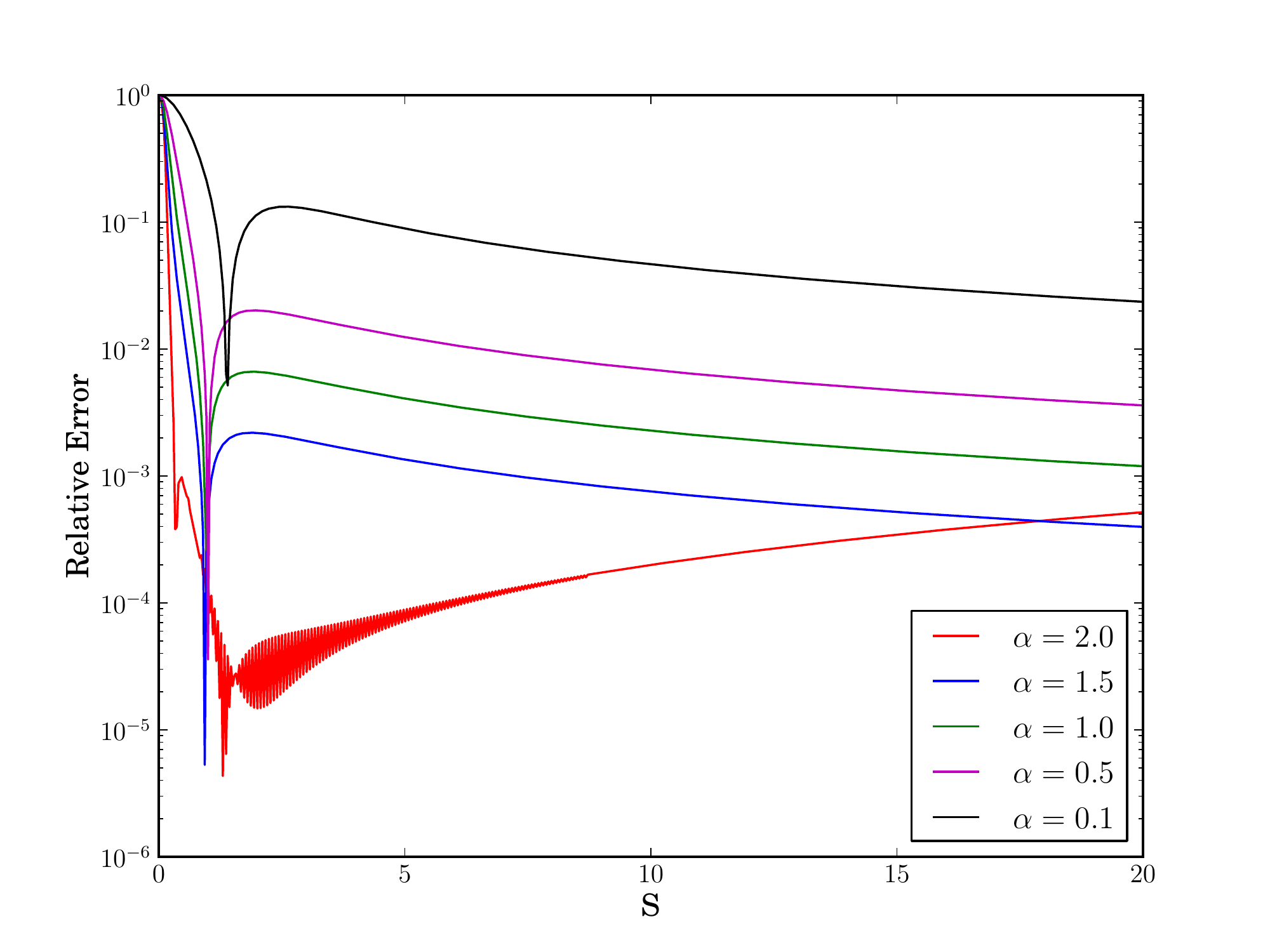}
 \caption{The relative error of our numerical method with respect to the analytic solution for a sharp $k$-space filter, Gaussian-random density field, and constant barrier. Results are shown for different values of the numerical parameter, $\alpha$. The mesh size is fixed at 600 in all cases.}
 \label{fig:RE1} 
\end{figure}

\begin{figure}
 \centering
 \includegraphics[width=12.0cm]{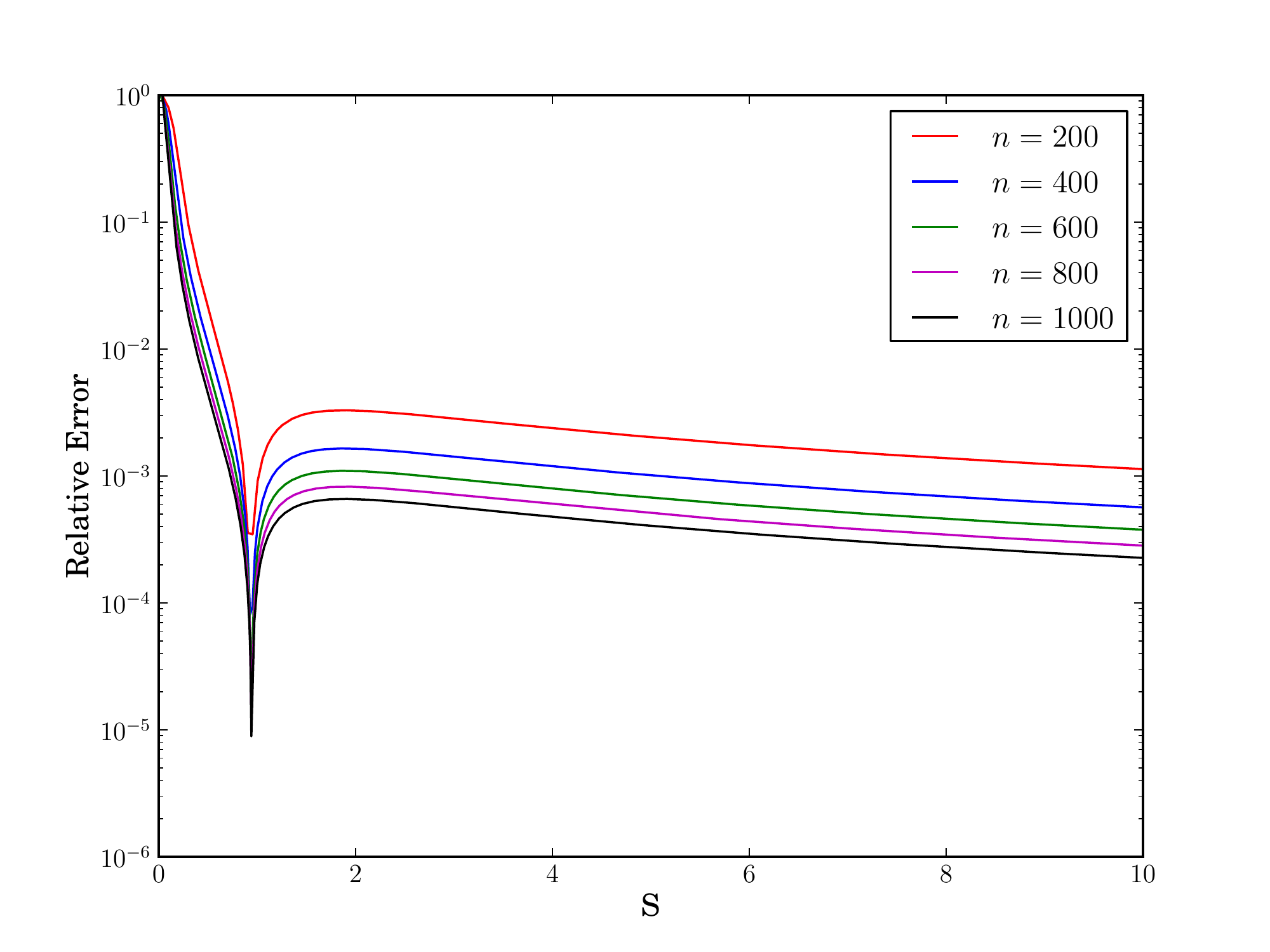}
 \caption{The relative error of our numerical method with respect to the analytic solution for a sharp $k$-space filter, Gaussian-random density field, and constant barrier.  Results are shown for different mesh sizes (as indicated in the figure; here $n$ is the number of mesh points in the range $S = [0,10]$), and for a fixed $\alpha = 1.5$.}
 \label{fig:RE2} 
\end{figure}

Figures~\ref{fig:RE1} and \ref{fig:RE2} illustrate the convergence achieved using this numerical recipe. In both cases we compare our numerical results with the analytic solution for a sharp $k$-space filter, a Gaussian random density field and a constant barrier. Figure~\ref{fig:RE1} shows the relative error in the numerical and analytic solution for different values of $\alpha$ and a fixed mesh size of 600. While the error is lowest for $\alpha=2$ over most of the range, it is clearly growing for the largest values of $S$. We find that it continues to do so as $S$ is increased further. As our solution is typically valid only for small $S$ this may not be a problem. Nevertheless, we find that adopting a value of $\alpha$ slightly less than $2$ results in acceptably small errors, and an error term that decreases with increasing $S$. Figure~\ref{fig:RE2} considers different mesh sizes for fixed $\alpha=1.5$. The numerical solution is stable (with increasing accuracy) as the mesh size is increased, demonstrating that our solutions are robust and numerically converged.

\section{Monte Carlo prescription} \label{app:MC}

In this section we briefly describe our Monte Carlo approach to computing the first crossing distribution. Our method uses eqn.~(\ref{eq:Pmemperturbationterms}) and derives the probability of trajectories originating from an arbitrary point $(\delta_{n},S_{n})$ to arrive at another arbitrary point $(\delta_{n+1},S_{n+1})$. Using our conventions of $A\equiv\langle\delta_{n}\delta_{n}\rangle$, $B\equiv\langle\delta_{n+1}\delta_{n+1}\rangle$, and $C\equiv\langle\delta_{n}\delta_{n+1}\rangle$, we find:

\begin{equation} \label{eq:Pmemperturbationterms22}
   P(\delta_{n};S_{n};\delta_{n+1};S_{n+1}) = \frac{1}{\sqrt{2\pi(B-C^2/A)}} \exp \left\{ -\frac{(\delta_{n+1}-C~\delta_{n} / A)^2}{2(B-C^2/A)}. \right\}
\end{equation} 

Now we can construct a set of trajectories which evolve through $S$, with the above equation giving the probability distribution for the next step in the walk. That is, at each step, assuming trajectory is located at point $(\delta_{n},S_{n})$ we assign $\delta_{n+1}$ by drawing from a Gaussian probability function with mean of $C\delta_n/A$ and variance of $(B-C^2/A)^{1/2}$ and randomly we choose a point to find its new location. We evolve each trajectory until it hits the barrier. Counting the number of trajectories crossing the barrier in range of $S$ and $S+\delta S$ and dividing it by $\delta S$ we find the probability density of trajectories crossing the barrier at that specific smoothing scale.

\section{Exact Solution Complexity} \label{app:Claim}

Our method for finding $f(S)$ in the case of a Gaussian random density field involves only one approximation. As described in footnote~\ref{fn3} this assumption is to ignore the correlated nature of random walks when evaluating $G(S,\delta,S^\prime,B_a(S^\prime))$ (eqn.~\ref{eq:IntegralEquationConditionGeneral}). This approximation cause our method to overestimate $f(S)$ for larger values of $S$, as can be seen in Figure~\ref{fig:ResultsMS12}.

We would therefore like to be able to find an expression for the conditional probability and unconditional probability distributions of trajectories, with the additional condition that the trajectories have never crossed the barrier prior to the point where we impose the first condition, say for $S < S_k$. We should thereforefind the following probability densities,

\begin{equation}
   P(\delta_0=0;\delta_k;S_k) =  \int_{-\infty}^{B(S_{1})} {\rm d} \delta_{1} \ldots \int_{-\infty}^{B(S_{k-1})} {\rm d} \delta_{k-1} W(\delta_0;\delta_1,\cdots,\delta_k;S_k).
\end{equation} 

and

\begin{equation}
   P(\delta_0=0;\delta_k;S_k;\delta_n;S_n) =  \int_{-\infty}^{B(S_{1})} {\rm d} \delta_{1} \ldots \int_{-\infty}^{B(S_{k-1})} {\rm d} \delta_{k-1}\int_{-\infty}^{+\infty} {\rm d} \delta_{k+1} \ldots \int_{-\infty}^{+\infty} {\rm d} \delta_{n-1} W(\delta_0;\delta_1,\cdots,\delta_k,\cdots,\delta_n;S_n).
\end{equation} 

To illustrate the difficulty in evaluating these probabilities we will consider just the first of these equations. First we offset all $\delta_i$'s to be zero at the barrier,

\begin{equation}
   P(\delta_0=0;\delta_k;S_k) =  \int_{-\infty}^{0} {\rm d} \delta_{1} \ldots \int_{-\infty}^{0} {\rm d} \delta_{k-1} W(\delta_0;\delta_1+B_a(S_i),\cdots,\delta_{k-1}+B_a(S_{k-1}),\delta_k;S_k).
\end{equation} 

Using the definition of the Heaviside step function we can then write the above equation as, 

\begin{align}
   P(\delta_0=0;\delta_k;S_k) = \int_{-\infty}^{+\infty} {\rm d} \delta_{1} \ldots \int_{-\infty}^{+\infty} {\rm d} \delta_{k-1} \int \mathcal{D}\lambda \exp{\left\{ -\frac{1}{2} A_{ij} \lambda_i \lambda_j + i \lambda_j (\delta_j + B_a(S_i)) + i \lambda_k \delta_k \right\}} \mathcal{H}(\delta_1) \cdots \mathcal{H}(\delta_{k-1}),
\end{align}

where $\mathcal{D}\lambda = {\rm d}\lambda_1/2\pi \cdots {\rm d}\lambda_k/2\pi$, $A_{ij} = \langle \delta_i\delta_j \rangle$, and $\mathcal{H}$ is the Heaviside step function. Taking the integral over all $\delta$'s gives,

\begin{equation}
   P(\delta_0=0;\delta_k;S_k) = \int \mathcal{D}\lambda~\exp \left\{ -\frac{1}{2} A_{ij} \lambda_i\lambda_j + i\lambda_j B_{a}(S_j) + i\lambda_k \delta_k  \right\} \left(\pi \delta_D(\lambda_1) + \frac{i}{\delta_1}\right) \cdots \left(\pi \delta_D(\lambda_{k-1}) + \frac{i}{\delta_{k-1}} \right),
\end{equation} 

where $A_{ij} = \langle \delta_i\delta_j \rangle$, and $\delta_D$ is the Dirac delta function. 

There is no analytic solution to this integral, and, furthermore, in the case of correlated random walks the inverse of matrix $A$ does not have an explicit form. If a solution to this equation were found, our method should give exact answers for all filters and barriers in the case of a Gaussian random field.

\end{document}